\documentclass{emulateapj}

\journalinfo{Published in the Astrophysical Journal}
\submitted{}

\begin{document}


\title{Relativistic Iron Line Emission from the Neutron Star Low-mass X-Ray Binary 4U~1636-536}

\shorttitle{Relativistic Iron Line Emission from 4U~1636-536}

\author{Dirk Pandel and Philip Kaaret}
\affil{Department of Physics and Astronomy, University of Iowa, Iowa City, IA 52242}
\author{Stephane Corbel}
\affil{Universit\'e Paris 7 Denis Diderot and Service d'Astrophysique, UMR AIM, CEA Saclay, F-91191 Gif sur Yvette, France}

\shortauthors{Pandel, Kaaret, and Corbel}


\begin{abstract}

We present an analysis of {\it XMM-Newton} and {\it RXTE} data from three observations
of the neutron star LMXB 4U~1636-536.
The X-ray spectra show clear evidence of a broad, asymmetric iron emission line
extending over the energy range 4--9 keV.
The line profile is consistent with relativistically broadened Fe~K$\alpha$ emission
from the inner accretion disk.
The Fe~K$\alpha$ line in 4U~1636-536 is considerably broader than the asymmetric
iron lines recently found in other neutron star LMXBs,
which indicates a high disk inclination.
We find evidence that the broad iron line feature is a combination of several K$\alpha$ lines
from iron in different ionization states.

\end{abstract}

\keywords{
accretion, accretion disks ---
binaries: close ---
stars: individual (4U~1636-536) ---
stars: neutron ---
X-rays: binaries ---
X-rays: stars
}


\section{INTRODUCTION}

Relativistically broadened, asymmetric Fe~K$\alpha$ lines from the inner accretion disk
have been observed in many supermassive and stellar-mass black holes
\citep[e.g.][]{2006AN....327..943F}.
In neutron star binaries, however, the iron lines are weaker,
and until recently observations did not clearly reveal a relativistic line profile.
\citet{2007ApJ...664L.103B} found an asymmetric Fe~K$\alpha$ line
in {\it XMM-Newton} spectra of the low-mass X-ray binary (LMXB) Serpens X-1
and showed that the line profile is consistent with fluorescent iron line
emission from the inner accretion disk.
Similar asymmetric Fe~K$\alpha$ lines were found by
\citet{2008ApJ...674..415C} in {\it Suzaku} spectra of the neutron star LMXBs
Serpens X-1, 4U~1820-30, and GX~349+2.
In this paper we present {\it XMM-Newton} and {\it RXTE} observations
of the neutron star LMXB 4U~1636-536.
We analyze the X-ray spectra to determine the profile of the relativistic Fe~K$\alpha$ line
and constrain the properties of the inner accretion disk.

4U~1636-536 (V801~Ara) is a well-studied, bursting LMXB consisting of a neutron star
in a 3.8~hr orbit with a 0.4 solar mass, 18th magnitude star \citep{1990A&A...234..181V}
and is located at a distance of $\sim$6~kpc \citep{2006ApJ...639.1033G}.
The X-ray timing properties of the binary have been studied extensively.
\citet{1996ApJ...469L..17Z} and \citet{1997ApJ...479L.141W} discovered
quasi-periodic oscillations (QPOs) at kHz frequencies.
The source also exhibits highly coherent burst oscillations at 581~Hz
which are likely related to the rotation of the neutron star
\citep{1997IAUC.6541....1Z,2002ApJ...577..337S}.
\citet{1999ApJ...514L..31K} found that the soft X-ray emission,
modulated at the kHz QPO frequency, lags behind the hard X-ray emission.
A possible explanation for this phase lag is the reprocessing of hard X-rays
in a cooler Comptonizing corona with a size of at most a few kilometers.
\citet{2007MNRAS.376.1139B} interpreted observations showing a
decline of the QPO coherence and rms amplitude at high QPO frequencies
as evidence that the inner radius of the accretion disk in 4U~1636-536
is usually larger than but sometimes approaches
the innermost stable circular orbit (ISCO).
However, \citet{2006MNRAS.371.1925M} argued that this decline is
not caused by effects related to the ISCO.

In this paper we report on three X-ray observations of 4U~1636-536 that were carried out
simultaneously with {\it XMM-Newton} and {\it RXTE}.
We present an analysis of the X-ray spectrum over the 0.5--100~keV energy range
and our results from modeling the continuum and relativistic Fe~K$\alpha$ line emission.
We use the Fe~K$\alpha$ line profile to derive constraints
on the disk inclination and inner disk radius.
We show that the line profile is not consistent with a single relativistic Fe~K$\alpha$ line
from a neutron star accretion disk
and that at least two lines from different ionization states of iron are needed
to adequately describe the line profile.
Finally, we discuss the implications of our findings for measurements of neutron star radii
based on Fe~K$\alpha$ line profiles.


\section{OBSERVATIONS}

4U~1636-536 was observed with {\it XMM-Newton} \citep{2001A&A...365L...1J}
on 2005 August 29 (observation ID 0303250201),
on 2007 September 28 (observation ID 0500350301),
and on 2008 February 27 (observation ID 0500350401).
The EPIC PN camera \citep{2001A&A...365L..18S} collected 30.0~ks of data
starting at 18:24~UT during the 2005 observation (hereafter observation 1),
30.6~ks of data starting at 15:45~UT during the 2007 observation (observation 2),
and 38.6~ks of data starting at 04:16~UT during the 2008 observation (observation 3).
The EPIC PN was operated in timing mode and with the medium blocking filter.
We processed the PN data from observations 1 and 2 with the {\it XMM-Newton}
SAS version 7.1.0 using the latest calibration files.
The data from observation 3 required special processing by the {\it XMM-Newton}
Science Operations Center because of the very large number of events.
We extracted source photons from the timing mode data using CCD rows RAWX~= 30--46
(29--45 for observation 3) and background photons using RAWX~= 2--18.
The full range of CCD columns RAWY was used.
We selected only events with PATTERN~$\le$~4 (singles and doubles)
and FLAG~=~0 and restricted our analysis to the energy range 0.5--12~keV.
The average PN source count rate in this energy range, excluding X-ray bursts,
was 230~s$^{-1}$ for observation 1, 470~s$^{-1}$ for observation 2,
and 620~s$^{-1}$ for observation 3.
These rates are below the 800~s$^{-1}$ timing mode pile-up limit
above which the spectral response would be deteriorated by photon pile-up
({\it XMM-Newton} User's Handbook).
The background count rate was less than 2\% of the source count rate
and did not exhibit significant flaring.
We find a strong line-like feature in the background spectrum near 0.45~keV.
We interpret this feature as the electronic noise peak for double events
that is also present for PN imaging modes but is shifted to higher energies in timing mode.

4U~1636-536 was observed with {\it RXTE} \citep{1993A&AS...97..355B}
simultaneously with each {\it XMM-Newton} observation.
The {\it RXTE} observations were somewhat longer than the {\it XMM-Newton} observations
and completely overlapped with the EPIC PN exposures.
No Earth occultation occurred during these observations
and the target was observed continuously with {\it RXTE}.
In this paper we consider only the data obtained with the HEXTE detector
\citep{1998ApJ...496..538R}.
We did not use the PCA data because of their lower energy resolution
compared to the EPIC PN data and because the PCA spectrum in the 3--12~keV
energy range showed an up to 30\% higher flux than the EPIC PN spectrum.
Similar excesses of the PCA flux compared to other instruments
have previously been reported \citep[e.g.][]{1999ApJ...512..892T,2003A&A...411L.343C}.
It was not possible to correct for the higher PCA flux by including a multiplicative
cross-calibration factor because the flux excess is strongly energy dependent
($\sim$30\% at 3~keV and $\sim$10\%--15\% at 10~keV).
We included in our analysis only those HEXTE data that were taken
simultaneously to the EPIC PN data.
The HEXTE data cover the entire duration of the EPIC PN exposures
with the exception of a single $<$1~ks data gap in each observation.
The HEXTE background rate and the telescope pointing were stable
during both observations.
We extracted HEXTE data for the energy range 20--100~keV
using the REX data analysis script.
For observation 1, we included in our analysis data from both HEXTE clusters.
For observations 2 and 3, we only used the data from cluster B because
cluster A was no longer able to obtain background measurements.


\section{SPECTRAL ANALYSIS}

\begin{deluxetable*}{llll}
\tablecaption{
Fits of Various XSPEC Models to the X-Ray Spectra of 4U~1636-536
\label{spectralmodels}}
\tablewidth{6.5in}
\tablehead{
\colhead{} & \colhead{Observation~1} & \colhead{Observation~2} & \colhead{Observation~3} \\
\colhead{Spectral Model} & \colhead{$\chi^2_\nu$ ($\chi^2$/dof)} & \colhead{$\chi^2_\nu$ ($\chi^2$/dof)}
                         & \colhead{$\chi^2_\nu$ ($\chi^2$/dof)}
}
\startdata
1. \texttt{vphabs*(compTT+diskbb)} \dotfill                             & 1.44 (172.5/120) & 1.80 (178.2/99)  & 1.28 (122.8/96)  \\
2. \texttt{vphabs*(compTT+diskbb+bbodyrad)} \dotfill                    & 1.36 (161.0/118) & 1.43 (139.0/97)  & 0.97 (91.4/94)   \\
3. \texttt{vphabs*(compTT+diskbb+bbodyrad+diskline)} \dotfill           & 1.23 (314.2/255) & 1.22 (285.3/234) & 1.18 (272.4/231) \\
4. \texttt{vphabs*(compTT+diskbb+bbodyrad+diskline+diskline)} \dotfill  & 1.23 (312.4/254) & 1.21 (281.1/233) & 1.19 (273.8/230)
\enddata
\tablecomments{
The $\chi^2$ values for fits of various XSPEC models to the EPIC PN and HEXTE spectra of 4U~1636-536.
Shown are the reduced $\chi^2$ ($\chi^2_\nu=\chi^2$/dof), the $\chi^2$,
and the number of degrees of freedom (dof).
For models 1 and 2 the energy range 3--10 keV was excluded from the fit.
}
\end{deluxetable*}

For the spectral analysis, we removed all X-ray bursts from the data
(three for observation 1 and one each for observations 2 and 3).
The EPIC PN spectra were binned at about 1/3 of the FWHM detector resolution
and were fitted simultaneously with the HEXTE data.
We did not include a multiplicative cross-calibration factor between the two instruments.
The spectral fitting was performed using XSPEC version 12.4 \citep{1996ASPC..101...17A}.
For the EPIC PN spectrum from observation 3, we find significant residuals between data and model
at $\sim$2.3~keV near the instrumental Au edge.
We attribute these residuals to a small offset in the photon energy determination
caused be a slightly inaccurate charge transfer efficiency (CTE) calibration
in timing mode \citep[see, e.g.,][]{2007A&A...461.1049S}.
We corrected for this energy offset by introducing a multiplicative gain correction factor
of 1.0020 which we determined from a fit.
This gain factor, which was fixed in our further spectral analysis,
completely removed the residuals near the instrumental Au edge.
No gain correction was necessary for the spectra from observations 1 and 2.

Initial fitting of the spectra from all three observations
showed significant residuals between data and model below 2~keV
near the K-shell absorption edges of O, Ne, Mg, and Si.
We attribute these residuals to an overly simplified description
of the interstellar absorption edges by current XSPEC models.
While these absorption models are generally adequate for EPIC PN spectra,
the data presented here has an exceptionally high signal-to-noise ratio
so that very small discrepancies between data and model become apparent.
We confirmed this interpretation with the RGS spectra which clearly show a complex structure
with several narrow absorption features near the K-shell absorption edges of oxygen and neon.
This edge structure is not adequately described by any of the absorption models in XSPEC
which model the edges as an exponential with a low-energy cutoff.
The structure of the oxygen and neon K edges in the RGS spectra is similar to the structure
found by \citet{2004ApJ...612..308J,2006ApJ...648.1066J} for several X-ray binaries
(including 4U~1636-536) and interpreted as absorption by neutral and ionized oxygen and neon
in the interstellar medium.
We find no emission lines in the RGS spectra that could explain the residuals
in the EPIC PN spectra.
Because we are mainly interested in the spectrum at higher energies near the Fe~K$\alpha$ line,
we simply excluded the data near the absorption edges with the largest residuals.
The following energy ranges were excluded from the fits:
$<$0.5~keV (O K~edge at 0.54~keV),
0.78--0.99~keV (Ne K~edge at 0.87~keV),
1.13--1.45~keV (Mg K~edge at 1.3~keV),
and 1.75--1.87~keV (Si K~edge at 1.8~keV).

In order to determine an appropriate model for the continuum emission,
we initially fitted the spectra while excluding the 3--10~keV energy range.
This excludes any contribution from a relativistically broadened,
fluorescent Fe~K$\alpha$ line which,
according to the \texttt{diskline} model in XSPEC \citep{1989MNRAS.238..729F},
is limited to the 3--10~keV energy range for accretion disks around neutron stars.
The X-ray spectra of LMXBs are typically characterized
by a hard component interpreted as Comptonization of soft photons by a hot corona,
a soft component thought to be blackbody radiation from the accretion disk,
and a broad Fe~K$\alpha$ line at 6.4~keV \citep[e.g.][]{2000ApJ...533..329B}.
To fit the X-ray spectrum of the continuum emission we therefore combined
a \texttt{compTT} component \citep{1994ApJ...434..570T} for a Comptonizing corona or boundary layer,
a \texttt{diskbb} disk blackbody component \citep[e.g.][]{1986ApJ...308..635M},
and a \texttt{vphabs} component for photoelectric absorption with variable abundances
(model 1).
In the \texttt{compTT} model we used the approximation for a disk geometry
and fixed the redshift at 0.
In the \texttt{vphabs} model we only varied the elemental abundances for O, Ne, Si, and Fe.
Abundances for the other elements were not well constrained by the fit
and were fixed at their solar values according to \citet{2000ApJ...542..914W}.
The spectra from the three observations are reasonably well fitted by model 1
with $\chi_\nu^2$ values of 1.44, 1.80, and 1.28 (Table~\ref{spectralmodels}).

We find that the fit can be significantly improved by adding a second blackbody component
(\texttt{bbodyrad}) with a temperature of $\sim$200~eV (model 2).
The additional component improves the $\chi_\nu^2$ for the three observations
to 1.36, 1.43, and 0.97, respectively.
A second blackbody component with a similar temperature was also introduced by
\citet{2006ApJ...651..416F} in their fitting of {\it BeppoSAX} spectra of 4U~1636-536
obtained during the high/soft state.
The region associated with the blackbody component has an apparent radius of $\sim$100~km,
which suggests that this emission originates from the accretion disk.
We note that the \texttt{diskbb} model \citep{1986ApJ...308..635M} was constructed
for accretion disks around black holes and may not correctly describe
the temperature profile in accretion disks around neutron stars
which are subjected to significant central illumination from the stellar surface.
The second blackbody component may represent deviations of the temperature profile
in the neutron star accretion disk from the \texttt{diskbb} model.
No further improvement in the fit was achieved by adding a third blackbody component
or other common spectral components.
The remaining discrepancy between data and model is probably caused
by small calibration uncertainties that are noticeable
because of the very high signal-to-noise ratio of the data
and, as mentioned above, by the overly simplified modeling of some absorption edges.

For our further analysis we included the data in the 3--10~keV energy range
which were previously excluded when fitting models 1 and 2.
The data in this energy range show a clear excess over the continuum flux
predicted by model 2 which is likely caused by relativistically broadened
Fe~K$\alpha$ line emission.
In order to fit the profile of this Fe~K$\alpha$ line,
we added to model 2 a \texttt{diskline} component \citep{1989MNRAS.238..729F}
which describes relativistically broadened line emission
from a neutron star accretion disk (model 3).
All parameters of the \texttt{diskline} model were allowed to vary during the fit.
We initially attempted to vary only the \texttt{diskline} parameters
and fix all other parameters at their values from model 2,
but we found that a better fit can be achieved by allowing the continuum model parameters
to vary as well.
This is an indication that the spectrum outside the 3--10~keV energy range is insufficient
to correctly determine all continuum model parameters.
Model 3 provides a good fit to the data with $\chi_\nu^2$ values of
1.23, 1.22, and 1.18, respectively.

\begin{figure}
\centering
\plotone{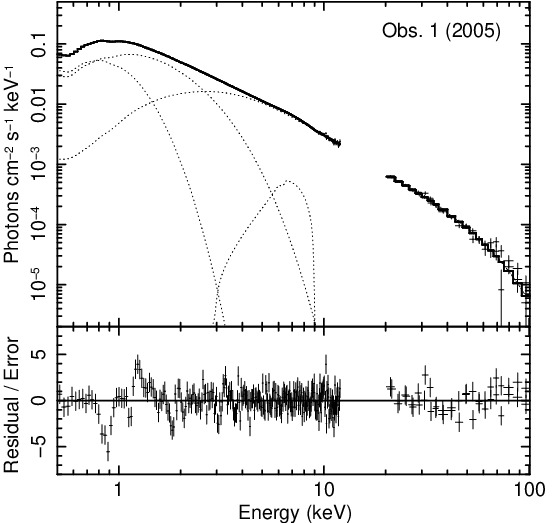}
\plotone{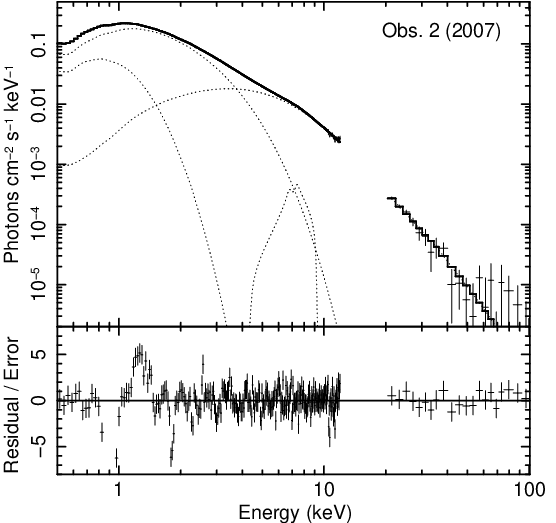}
\plotone{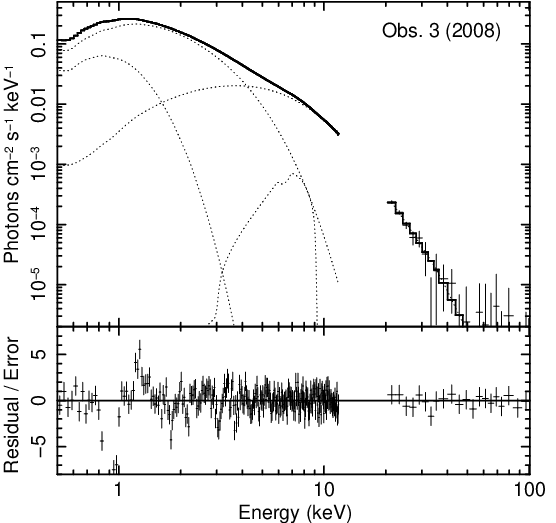}
\caption{
EPIC PN and HEXTE spectra of 4U~1636-536 for the three observations.
The top panels show the observed flux ({\it error bars}),
the best fit with model 3 ({\it solid line}),
and the individual additive model components ({\it dotted lines}).
The individual model components are, from left to right,
\texttt{bbodyrad}, \texttt{diskbb}, \texttt{compTT}, and \texttt{diskline}.
The bottom panels show the residuals between the observed flux and the model
divided by the 1~$\sigma$ error of each data point.
}
\label{spec1}
\end{figure}

\begin{figure}
\centering
\plotone{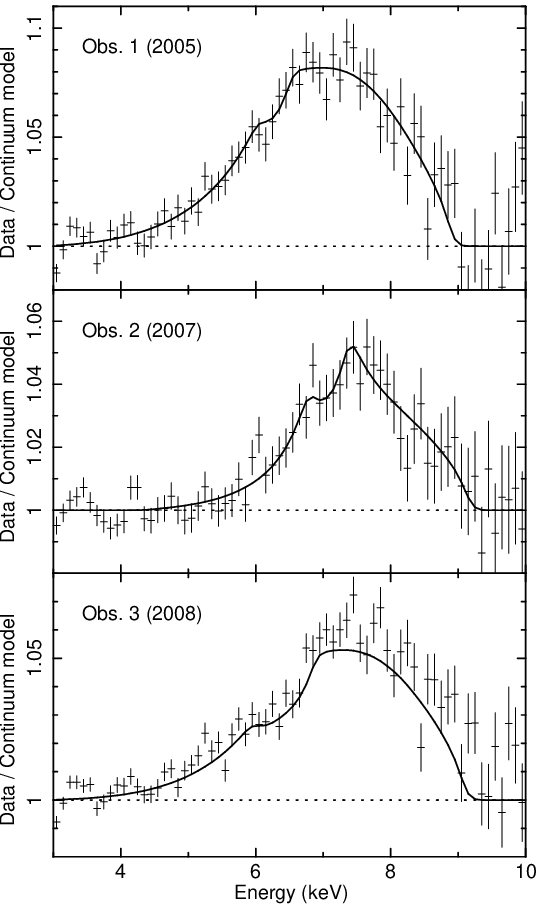}
\caption{
Relativistic iron line profiles in 4U~1636-536.
Shown are the observed flux ({\it error bars}) and the model flux ({\it solid line}),
normalized to the continuum flux from model 3
(i.e.\ the model flux without the \texttt{diskline} component).
}
\label{spec2}
\end{figure}

Figure \ref{spec1} shows the observed spectra for the three observations,
the best fit with model 3, the individual model components,
and the residuals between data and model.
The excess of the line emission over the continuum model
is shown in Figure \ref{spec2}.
The figure clearly shows a broad and asymmetric Fe~K$\alpha$ line
with a peak flux of 5\%--8\% over the continuum.
The line profile is well described by the \texttt{diskline} model.
The best-fit parameters of the continuum and line components for model 3
are shown in Tables~\ref{contpars} and~\ref{linepars}.
Also shown are the uncertainties of the parameters at a 90\% confidence level
(95\% for upper and lower limits).
When calculating the confidence regions, all continuum and line parameters were allowed to vary.
We found that the commonly used \texttt{error} and \texttt{steppar} commands in XSPEC
frequently failed to converge or underestimated the uncertainties.
We therefore calculated the confidence regions for many of the parameters
by manually searching for the parameter value that produced
the appropriate change in $\chi^2$.

The continuum parameters for model 3 are generally consistent between the three observations,
although the fit indicates a higher disk temperature and stronger Comptonization component
for the later observations.
We find a significant difference in the rest-frame energy $E_0$ of the Fe~K$\alpha$ line.
For observations 1 and 3, $E_0$ is consistent with 6.4~keV for weakly ionized iron,
whereas, for observation 2, $E_0$ is close to 7.0~keV for highly ionized Fe~{\sc xxvi}.
The power-law index $\beta$ of the emissivity profile is in the typical range
found for iron lines from black hole accretion disks.
The \texttt{diskline} model provides lower limits on the disk inclination
of $81^\circ$ for observations 1 and 3 and $64^\circ$ for observation 2.
A lower limit of $81^\circ$ is clearly inconsistent with the fact
that 4U~1636-536 is not an eclipsing system
as well as with the results by \citet{2006MNRAS.373.1235C}, who constrained
the orbital inclination to 36$^\circ$--74$^\circ$ using phase-resolved spectroscopy.
The high limit on the disk inclination for observations 1 and 3 is likely the result
of a slightly broader line profile compared to observation 2.
Our fit with the \texttt{diskline} model suggests that the line profile is broader
than physically possible for a single Fe~K$\alpha$ emission line.
One possible explanation for a broader line profile is a smaller radius
of the inner disk edge $R_{\rm in}$.
However, the best-fit value of $R_{\rm in}$ for observations 1 and 3 is already
at $6R_g$ ($R_g=GM/c^2$), the lower limit of the \texttt{diskline} model
and the radius of the innermost stable circular orbit (ISCO) for a nonrotating
neutron star.
An inner disk radius much smaller than $6R_g$, while possible for rotating black holes,
is unlikely for the accretion disk around the neutron star in 4U~1636-536,
which is rotating at a rate of 581 Hz.
The most likely explanation for a broader line profile is the presence of
several blended Fe~K$\alpha$ lines with different rest-frame energies.
The presence of iron in more than one ionization state is already indicated by the difference
in $E_0$ between the three observations (6.30 and 6.43~keV vs.\ 7.06~keV).

To test whether blended Fe~K$\alpha$ lines are a viable explanation,
we added a second \texttt{diskline} component to the model (model 4).
Because the iron lines are broad and significantly blended,
the parameters of the two \texttt{diskline} components are strongly correlated,
and it is not possible to fit all line parameters independently.
We therefore fixed the line energy of the \texttt{diskline} components at 6.4 and 7.0~keV,
respectively, and linked the disk inclination parameters.
We also found that the outer disk radius $R_{\rm out}$ is poorly constrained by the fit,
so we kept it fixed at 1000~$R_g$.
The continuum parameters, which were allowed to vary simultaneously with the line parameters,
did not change significantly compared to model 3.
Adding the second line component does not significantly improve the fit (Table~\ref{spectralmodels}),
but it does result in more reasonable line parameters (Table~\ref{linepars}).
The lower limits on the disk inclination change to 64$^\circ$ and 65$^\circ$,
respectively, which, unlike for model 3,
are now consistent with the 36$^\circ$--74$^\circ$ constraint by \citet{2006MNRAS.373.1235C}.
The inner radius $R_{\rm in}$ for the 7.0~keV component appears to be generally smaller
than for the 6.4~keV component as is expected for an accretion disk
in which the temperature increases towards the center.
The flux ratio of the 7.0~keV and the 6.4~keV component increases from 0.9 to 1.6 to 4.9
between the three observations, indicating that iron is generally in a higher ionization state
for the later observations.
This is consistent with the increase in X-ray flux between observations.
We conclude that the presence of iron at different ionization states is a viable explanation
for the unusually broad line profiles and that at least two blended Fe~K$\alpha$ lines
are necessary to obtain consistent line parameters.

\begin{deluxetable*}{lccc}
\tablecaption{
Fit Parameters of the Continuum Components of Model 3
\label{contpars}}
\tablewidth{7.0in}
\tablehead{
\colhead{Parameter}               & \colhead{Observation~1}  & \colhead{Observation~2}  & \colhead{Observation~3}
}
\startdata
\\[-3ex]
\multicolumn{4}{c}{\texttt{vphabs}\tablenotemark{a}} \\
\hline\\[-2ex]
$N_{\rm H}$ ($10^{21}$~cm$^{-2}$) \dotfill  & $3.63\pm0.12$      & $3.83\pm0.10$      & $3.83\pm0.08$      \\
O abundance \dotfill                                & $1.37\pm0.04$      & $1.27\pm0.03$      & $1.27\pm0.03$      \\
Ne abundance \dotfill                               & $1.7\pm0.2$        & $1.2\pm0.2$        & $1.3\pm0.2$        \\
Si abundance \dotfill                               & $1.7\pm0.4$        & $2.0\pm0.4$        & $0.8\pm0.4$        \\
Fe abundance \dotfill                               & $1.49\pm0.10$      & $1.47\pm0.11$      & $1.48\pm0.09$      \\
\hline\\[-2ex]
\multicolumn{4}{c}{\texttt{compTT}\tablenotemark{b}} \\
\hline\\[-2ex]
$T_0$ (keV) \dotfill                              & $0.98\pm0.02$      & $1.45\pm0.02$      & $1.68\pm0.03$      \\
$T_{\rm C}$ (keV) \dotfill                        & $14.2\pm0.9$\phn   & $>$9\phd\phn\phs   & $>$4\phd\phn\phs   \\
$\tau_{\rm C}$ \dotfill                           & $2.34\pm0.12$      & $<$1.3\phs         & $<$2.7\phs         \\
Normalization ($10^{-3}$) \dotfill                & $9.4\pm0.7$        & $9\pm8$            & $25\pm22$          \\
Flux ($10^{-10}$~erg~cm$^{-2}$~s$^{-1}$) \dotfill & 6.1                & 8.2                & 10.0\phn           \\
\hline\\[-2ex]
\multicolumn{4}{c}{\texttt{diskbb}\tablenotemark{c}} \\
\hline\\[-2ex]
$T_{\rm disk}$ (eV) \dotfill                      & $640\pm30$\phn     & $780\pm20$\phn     & $900\pm20$\phn     \\
Normalization \dotfill                            & $150\pm30$\phn     & $200\pm13$\phn     & $152\pm7$\phn\phn  \\
$R_{\rm disk} \sqrt{\cos{i}}$ (km) \dotfill       & $7.3\pm0.7$        & $8.5\pm0.3$        & $7.4\pm0.2$        \\
Flux ($10^{-10}$~erg~cm$^{-2}$~s$^{-1}$) \dotfill & 4.5                & 13.5\phn           & 18.3\phn           \\
\hline\\[-2ex]
\multicolumn{4}{c}{\texttt{bbodyrad}\tablenotemark{d}} \\
\hline\\[-2ex]
$T_{\rm bb}$ (eV) \dotfill                        & $174\pm9$\phn\phn  & $177\pm7$\phn\phn  & $193\pm6$\phn\phn  \\
Normalization ($10^{3}$) \dotfill                 & $29\pm6$\phn       & $29\pm8$\phn       & $23\pm5$\phn       \\
$R_{\rm bb}$ (km) \dotfill                        & $102\pm11$\phn     & $102\pm14$\phn     & $91\pm10$          \\
Flux ($10^{-10}$~erg~cm$^{-2}$~s$^{-1}$) \dotfill & 1.9                & 2.0                & 2.3
\enddata
\tablecomments{
Uncertainties are given at a 90\% confidence level
and limits at a 95\% confidence level.
Fluxes for individual model components are
unabsorbed and for the energy range 0.5--10 keV.
$R_{\rm disk} \sqrt{\cos{i}}$ and $R_{\rm bb}$ were calculated from the model normalizations
assuming a distance of 6~kpc \citep{2006ApJ...639.1033G}.
}
\tablenotetext{a}{
Photoelectric absorption model with variable abundances.
$N_{\rm H}$ is the neutral hydrogen column density.
Elemental abundances are given relative to their solar values
according to \citet{2000ApJ...542..914W}.
Abundances for elements not shown were fixed at their solar values.
}
\tablenotetext{b}{
Comptonization model according to \citet{1994ApJ...434..570T}.
$T_0$ is the temperature of the soft seed photons,
$T_{\rm C}$ the temperature of the Comptonizing plasma,
and $\tau_{\rm C}$ the optical depth.
A disk geometry was assumed for the model,
and the redshift was fixed at 0.
}
\tablenotetext{c}{
Disk-blackbody model \citep[e.g.][]{1986ApJ...308..635M}.
$T_{\rm disk}$ is the temperature at the inner disk edge,
$R_{\rm disk}$ the apparent inner disk radius \citep[see][]{1998PASJ...50..667K},
and $i$ the disk inclination.
}
\tablenotetext{d}{
Blackbody model.
$T_{\rm bb}$ is the blackbody temperature and
$R_{\rm bb}$ the apparent radius of the emitting region.
}
\end{deluxetable*}

\begin{deluxetable*}{lcccccc}
\tablecaption{
Fit Parameters of the Iron Line Components of Models 3 and 4
\label{linepars}}
\tablewidth{7.0in}
\tablehead{
\multicolumn{7}{c}{Model 3} \\
\colhead{Parameter} & \multicolumn{2}{c}{Observation~1} & \multicolumn{2}{c}{Observation~2} & \multicolumn{2}{c}{Observation~3}
}
\startdata
$E_0$ (keV) \dotfill
			& \multicolumn{2}{c}{$6.30\pm0.09$}
			& \multicolumn{2}{c}{$7.06\pm0.10$}
			& \multicolumn{2}{c}{$6.43\pm0.09$}  \\
$\beta$ \dotfill
			& \multicolumn{2}{c}{$-2.72\pm0.11$\phs}
			& \multicolumn{2}{c}{$-2.32\pm0.22$\phs}
			& \multicolumn{2}{c}{$-2.73\pm0.13$\phs}  \\
$R_{\rm in}$ ($GM/c^2$) \dotfill
			& \multicolumn{2}{c}{$<$6.3\phs}
			& \multicolumn{2}{c}{$<$13.3\phs}
			& \multicolumn{2}{c}{$<$6.3\phs}  \\
$R_{\rm out}$ ($GM/c^2$) \dotfill
			& \multicolumn{2}{c}{$820^{+2900}_{-440}$}
			& \multicolumn{2}{c}{$690^{+630}_{-400}$}
			& \multicolumn{2}{c}{$210^{+180}_{-70}$}  \\
$i$ (deg) \dotfill
			& \multicolumn{2}{c}{$>$81\phs}
			& \multicolumn{2}{c}{$>$64\phs}
			& \multicolumn{2}{c}{$>$81\phs}  \\
Normalization ($10^{-3}$) \dotfill
			& \multicolumn{2}{c}{$1.49\pm0.13$}
			& \multicolumn{2}{c}{$0.91\pm0.20$}
			& \multicolumn{2}{c}{$2.00\pm0.20$}  \\
Equivalent width (eV) \dotfill
			& \multicolumn{2}{c}{215}
			& \multicolumn{2}{c}{98}
			& \multicolumn{2}{c}{140}  \\
Flux ($10^{-12}$~erg~cm$^{-2}$~s$^{-1}$) \dotfill
			& \multicolumn{2}{c}{15.7}
			& \multicolumn{2}{c}{10.2}
			& \multicolumn{2}{c}{21.5}  \\[0.5ex]
\hline\\[-2ex]
\multicolumn{7}{c}{Model 4} \\
\colhead{Parameter} & \multicolumn{2}{c}{Observation~1} & \multicolumn{2}{c}{Observation~2} & \multicolumn{2}{c}{Observation~3} \\
 & Line 1 & Line 2 & Line 1 & Line 2 & Line 1 & Line 2  \\
\hline\\[-2ex]
$E_0$ (keV) \dotfill
			& (6.4)				& (7.0)
			& (6.4)				& (7.0)
			& (6.4)				& (7.0)  \\
$\beta$ \dotfill
			& $-2.7\pm0.2$			& $<$--2.7
			& $<$--2.6			& $-2.7^{+0.3}_{-1.6}$
			& $<$--2.7			& $-2.7\pm0.2$  \\
$R_{\rm in}$ ($GM/c^2$) \dotfill
			& $<$17.8			& $<$11.9
			& $9.3^{+5.6}_{-2.4}$		& $<$9.8
			& $<$33.6			& $<$10.9  \\
$R_{\rm out}$ ($GM/c^2$) \dotfill
			& (1000)			& (1000)
			& (1000)			& (1000)
			& (1000)			& (1000)  \\
$i$ (deg) \dotfill
			& $>$64				& $>$64
			& $>$65				& $>$65
			& $>$64				& $>$64  \\
Normalization ($10^{-3}$) \dotfill
			& $0.7^{+0.3}_{-0.2}$		& $0.6^{+0.2}_{-0.4}$
			& $0.5^{+0.6}_{-0.3}$		& $0.8^{+0.2}_{-0.4}$
			& $0.3\pm0.2$			& $1.3\pm0.2$  \\
Equivalent width (eV) \dotfill
			& 108				& 131
			& 54				& 84
			& 23				& 100  \\
Flux ($10^{-12}$~erg~cm$^{-2}$~s$^{-1}$) \dotfill
			& 7.7				& 7.0
			& 5.7				& 8.9
			& 3.0				& 14.6
\enddata
\tablecomments{
$E_0$ is the rest-frame energy of the iron emission line,
$\beta$ the power-law index of the emissivity dependence on radius,
$R_{\rm in}$ and $R_{\rm out}$ the inner and outer disk radius
(in units of $GM/c^2$ with $M$ being the neutron star mass),
and $i$ the disk inclination.
Also shown is the equivalent line width and the integrated flux
for each \texttt{diskline} component.
Uncertainties are given at a 90\% confidence level
and limits at a 95\% confidence level.
Values in parenthesis indicate that the parameter was fixed during the fit.
}
\end{deluxetable*}


\section{DISCUSSION}

We have analyzed X-ray spectra of the neutron star LMXB 4U~1636-536
obtained with {\it XMM-Newton} and {\it RXTE} in 2005, 2007, and 2008.
The very high signal-to-noise ratio of the spectra allowed us to clearly detect
a broad, relativistic Fe~K$\alpha$ line from the inner accretion disk.
The line is significantly broader than the asymmetric Fe~K$\alpha$ lines
recently found in other neutron star LMXBs \citep{2007ApJ...664L.103B,2008ApJ...674..415C}.
The broader line profile is likely the result of a high disk inclination
in 4U~1636-536.
The inclination angles derived from iron line profiles in other neutron star LMXBs
have so far been comparatively low.
As pointed out by \citet{2008ApJ...674..415C}, this is likely a selection effect
because the narrower lines in low-inclination systems are more easily detectable.
With the high signal-to-noise ratio of the 4U~1636-536 spectra,
we have now been able to measure the broader line profile
in a high inclination LMXB.
Our analysis of the Fe~K$\alpha$ line profile places a lower limit of $64^\circ$
on the disk inclination in 4U~1636-536.
This limit is consistent with the 36$^\circ$--74$^\circ$ constraint on the orbital inclination
by \citet{2006MNRAS.373.1235C} and with the nondetection of eclipses.

When fitting the iron line profile with a single relativistic line component,
we find a significant difference in the rest-frame line energy between the three observations.
This difference is likely caused by a change in the ionization profile of the disk
related to the change in X-ray luminosity between the three observations.
The line energy derived for the second observation is close to 6.97~keV,
the K$\alpha$ line energy of Fe~{\sc xxvi}, which suggests that Fe~{\sc xxvi}
contributes significantly to the observed line profile.
According to our fit with a disk-blackbody model,
the highest temperature in the disk is $\sim$0.9~keV.
Because plasma in thermal equilibrium at this temperature does not contain
a significant fraction of Fe~{\sc xxvi}, it is evident that the ionization profile in the disk
is strongly affected by photoionization.

We find that the Fe~K$\alpha$ line profile for two of the observations is too broad
to be adequately described by a single relativistic emission line
with physically reasonable values of disk inclination and inner disk radius.
The broader than expected line profile can be explained by
overlapping K$\alpha$ lines from iron in different ionization states.
The presence of multiple ionization states is also indicated by the difference
in the fitted line energy between the three observations.
It is evident that multiple line components need to be considered to adequately model
the relativistic iron line profiles in neutron star LMXBs.
We obtained reasonable line parameters when fitting two iron lines
with different rest-frame energies.
This is obviously an oversimplification, since many ionization states are probably
contributing to the iron line profile.
However, because the lines are broad and overlap considerably,
it is not possible to constrain the parameters of all line components independently.
Even with only two line components, the fit parameters are already strongly correlated,
leading to larger parameter uncertainties than a fit with a single line component.
In order to adequately model the contribution from multiple ionization states
when fitting relativistic Fe~K$\alpha$ line profiles, improved models are needed that can predict
the ionization profile in the accretion disk.

The Fe~K$\alpha$ line profiles in neutron star LMXBs can in principle be used
to place upper limits on the neutron star radius by constraining the inner disk radius
\citep{2008ApJ...674..415C}.
Previously, these line profiles have been fitted with only a single line component.
However, if more than one ionization state contributes significantly to the Fe~K$\alpha$ emission,
the line profile will be broader than for a single relativistic line,
and a fit with a single line model may underestimate the inner disk radius
and thus the limit on the neutron star radius.
This can be clearly seen for two of our observations for which the upper limit on $R_{\rm in}$
increases from $6.3R_g$ to $11.9R_g$ and $10.9R_g$, respectively,
when the iron line profile is fitted with two line components instead of a single line component.
In contrast, the upper limit on $R_{\rm in}$ for the second observation
decreases from $13.3R_g$ to $9.8R_g$.
It is evident that the constraints on the inner disk radius strongly depend
on the assumptions made about the contributing ionization states of iron
and that a better understanding of the ionization profile in the disk is needed
to obtain reliable limits on the neutron star radius.
The contribution of multiple ionization states may also be important for the interpretation
of some iron line profiles in accreting black holes.
We note that the disk inclination and inner disk radius are strongly anti-correlated
when fitting broad iron line profiles, which can lead to large uncertainties
of the two parameters.
Prior knowledge of the disk inclination can significantly reduce the uncertainty
of the inner disk radius.

It was shown by \citet{2007ApJ...656.1056L} that broad iron line features
can also be produced by Compton scattering of line photons in a strong outflow.
The expected line profiles for this process are generally characterized
by a narrow line at $\sim$6.4--6.6~keV from fluorescence in the outflow
and a broad, redshifted component below 7~keV from downscattering of the line photons.
These line profiles differ qualitatively from those found in 4U~1636-536
which show significant emission above 7~keV and no narrow line component (Fig.~\ref{spec2}).
We also note that Compton scattering in an outflow only contributes significantly
to the iron line emission if the optical depth is at least of order unity,
which requires a mass outflow rate on the order of the Eddington mass accretion rate.
The X-ray luminosity we observed in 4U~1636-536 was only $\sim$5\% of the Eddington luminosity,
which suggests that the rate of any outflow in 4U~1636-536 was significantly
below the Eddington mass accretion rate.
It therefore seems unlikely that a large fraction of the observed iron line emission
was produced in an outflow.


\acknowledgments

The authors thank Sudip Bhattacharyya for helpful discussions.
D.~P.\ and P.~K.\ acknowledge partial support from NASA grants NNG05GQ18G and NNX07AV06G.
This work is based on observations obtained with {\it XMM-Newton}, an ESA science mission
with instruments and contributions directly funded by ESA member states and the USA (NASA).

{\it Facilities:} \facility{XMM, RXTE}


\bibliographystyle{apj}
\bibliography{4u1636}

\end{document}